\documentclass[preprintpaper]{pnastwo-preprint}

\usepackage{graphicx}
\usepackage{amssymb,amsfonts,amsmath}

\url{ www.pnas.org/cgi/doi/10.1073/pnas.0806887106 }
\copyrightyear{2008}
\issuedate{Issue Date}
\volume{Volume}
\issuenumber{Issue Number}

\usepackage{cite}

\makeatletter
\def\@cite#1#2{$(#1\if@tempswa , #2\fi)$}
\makeatother

\begin{document}

\title{Nonlinear threshold behavior during the loss of Arctic sea ice}

\author{I. Eisenman\affil{1}{Department of Earth \& Planetary Sciences, Harvard University, Cambridge, Massachusetts, USA}$^{,1,2}$ and J.S. Wettlaufer\affil{2}{Department of Geology \& Geophysics and Department of Physics, Yale University, New Haven, Connecticut, USA, and Nordic Institute for Theoretical Physics, Roslagstullsbacken 23, SE-106 91 Stockholm, Sweden}}

\contributor{Proceedings of the National Academy of Sciences
of the United States of America {\bf (in press, 2008)}}

\maketitle

\begin{article}


\begin{abstract}

In light of the rapid recent retreat of Arctic sea ice, a number of studies have discussed the possibility of a critical threshold (or ``tipping point'') beyond which the ice-albedo feedback causes the ice cover to melt away in an irreversible process. The focus has typically been centered on the annual minimum (September) ice cover, which is often seen as particularly susceptible to destabilization by the ice-albedo feedback. Here we examine the central physical processes associated with the transition from ice--covered to ice--free Arctic Ocean conditions. We show that while the ice-albedo feedback promotes the existence of multiple ice cover states, the stabilizing thermodynamic effects of sea ice mitigate this when the Arctic Ocean is ice-covered during a sufficiently large fraction of the year. These results suggest that critical threshold behavior is unlikely during the approach from current perennial sea ice conditions to seasonally ice-free conditions. In a further warmed climate, however, we find that a critical threshold associated with the sudden loss of the remaining wintertime-only sea ice cover may be likely.


\end{abstract}



\dropcap{T}he retreat of Arctic sea ice during recent decades \cite{Stroeve-Serreze-Fetterer-et-al-2005:tracking} is believed to be augmented by the difference in albedo (i.e., reflectivity) between sea ice and exposed ocean waters \cite{Perovich-Light-Eicken-et-al-2007:increasing}.
Because bare or snow-covered sea ice is highly reflective to solar radiation, the increasing area of open water that is exposed as sea ice recedes leads to an increase in absorbed solar radiation, thereby contributing to further ice retreat. A number of recent studies have discussed the possibility that this positive ice-albedo feedback will cause the rapidly declining annual minimum (September) sea ice cover to cross a critical threshold, after which the sea ice will melt back on an irreversible trajectory to a seasonally ice-free state \cite{Lindsay-Zhang-2005:thinning, Overpeck-Sturm-Francis-et-al-2005:arctic, Serreze-Francis-2006:arctic, Holland-Bitz-Tremblay-2006:future, 
Maslanik-Stroeve-Drobot-et-al-2007:younger, Lenton-Held-Kriegler-et-al-2008:tipping, 
Merryfield-Holland-Monahan-2006:multiple}.

Heuristically, one might expect in a simple annual mean picture of the Arctic Ocean that completely ice-covered and ice-free stable states could co-exist under the same climate forcing. The ice-free state would remain warm due to the absorption of most incident solar radiation, whereas the ice-covered state would reflect most solar radiation and remain below the freezing temperature. In such a picture, these two stable states would be separated by an unstable intermediate state in which the Arctic Ocean is partially covered by ice and absorbs just enough solar radiation such that it remains at the freezing temperature: adding a small amount of additional sea ice to this unstable state would lead to less solar absorption, cooling, and a further extended sea ice cover. If the background climate warmed, the unstable state would require an increased ice extent to reflect sufficient solar radiation to remain at the freezing temperature.  Beyond a critical threshold, the background climate would become so warm that the ice-covered state would reach the freezing temperature. At this point the stable ice-covered state and unstable intermediate state would merge and disappear in a saddle-node bifurcation, leaving only the warm ice-free state \cite{Budyko-1969:effect, Sellers-1969:global, North-1990:multiple}. This scenario suggests that if an ice-covered Arctic Ocean were warmed beyond the bifurcation point, there would be a rapid transition to the ice-free state. It would be an irreversible process in the sense that the Arctic Ocean would refreeze only after the climate had cooled to a second bifurcation point at which even an ice-free Arctic Ocean would become sufficiently cold to freeze, representing a significantly colder background climate than the original point at which the ice disappeared. Thus the ice-albedo feedback could, in principle, cause a hysteresis loop in the Arctic climate response to warming.

Here we investigate the central physical processes underlying the possibility of such a bifurcation threshold in future sea ice retreat. We illustrate the discussion with a seasonally varying model of the Arctic sea ice--ocean--atmosphere climate system.

\section{Arctic sea ice and climate model}

The theory presented here describes the thermal evolution of sea ice, ocean mixed layer, and an energy balance atmosphere that is in steady-state with the underlying surface forcing, including also representations of dynamic sea ice export and diffusive atmospheric meridional heat transport. The sea ice thermodynamics in this model is an approximation of the full heat conduction equation of Maykut \& Untersteiner \cite{Maykut-Untersteiner-1971:some}, which provides the thermodynamic basis for most current sea ice models. Ice grows during the winter at the base, and when the surface reaches the freezing temperature in summer, ablation occurs at the surface as well as at the base. This model produces an observationally consistent simulation of the modern Arctic sea ice seasonal cycle using a single one-dimensional nonautonomous ordinary differential equation with observationally-based seasonally-varying parameters. Here we provide a brief summary of the model equations, which are fully derived from basic physical principles in the {\it Supporting Information}.
  
The state variable $E$ represents the energy per unit area stored in sea ice as latent heat when the ocean is ice-covered or in the ocean mixed layer as sensible heat when the ocean is ice-free,
\begin{equation}
E\equiv \begin{cases}-L_i h_i & E<0 \textrm{   [sea ice]}\\
c_{ml}H_{ml}T_{ml} & E \ge 0 \textrm{   [ocean]} \end{cases},
\label{eq:sum0}
\end{equation}
where $L_i$ is the latent heat of fusion for sea ice, $h_i$ is the sea ice thickness, $c_{ml}$ is the mixed layer specific heat capacity, and $H_{ml}$ is the mixed layer depth. The ocean mixed layer temperature is written in terms of departure from the freezing point, $T_{ml} \equiv \tilde{T}_{ml}-\tilde{T}_{fr}$, where $\tilde{T}_{ml}$ is the ocean mixed layer temperature and $\tilde{T}_{fr}$ is taken to be 0\mbox{$^\circ$}C. The time evolution of $E$ is proportional to the net energy flux,
\begin{eqnarray}
\frac{d E}{d t} &=& \left[1-\alpha(E)\right] F_S(t) -F_0(t) + \Delta F_0 \nonumber\\
 && - F_T(t) T(t,E)+F_B + v_0 \mathcal{R}(-E) .
\label{eq:sum1}
\end{eqnarray}
Here the Stefan-Boltzmann equation for outgoing longwave radiation has been linearized in the surface temperature departure from the freezing point, $T(t,E) \equiv \tilde{T}(t,E)-\tilde{T}_{fr}$, as $F_0(t)+F_T(t)T(t,E)$, where the parameters also account for the effects of a partially opaque atmosphere and atmospheric heat flux convergence which is a function the meridional temperature gradient. The seasonally varying values of $F_0(t)$ and $F_T(t)$ are derived using an atmospheric model that incorporates observations of Arctic cloudiness \cite{Maykut-Church-1973:radiation}, surface air temperature south of the Arctic \cite{Kalnay-Kanamitsu-Kistler-et-al-1996:ncep}, and atmospheric transport into the Arctic \cite{Nakamura-Oort-1988:atmospheric}.  

The term $\Delta F_0$ represents a specified perturbation to the surface heat flux, which is zero by default but can be increased to prescribe a warming in the model. Incident surface shortwave radiation $F_S(t)$
and basal heat flux $F_B$ are specified at central Arctic values \cite{Maykut-Untersteiner-1971:some}. The final term in equation \eqref{eq:sum1} accounts for an observationally-based constant ice export of $v_0=$10\% yr$^{-1}$ \cite{Kwok-Cunningham-Pang-2004:fram} when ice is present ($E<0$), with the ramp function $\mathcal{R}(x)$ defined to equal $x$ when $x\ge 0$ and 0 when $x<0$. 

The surface temperature $T(t,E)$ can evolve between three different regimes. {\it (i)} When ice is present ($E<0$) and the surface temperature is below the freezing point ($T(t,E)<0$), it is calculated from a balance between the heat flux above the ice surface and upward heat flux in the ice, $-\left[1-\alpha(E)\right] F_S(t) +F_0(t)- \Delta F_0 + F_T(t) T(t,E)=-k_i T(t,E)/h_i=k_i L_i T(t,E)/E$. {\it (ii)} When the surface temperature warms to the freezing point ($T(t,E)=0$), it remains at this point while the ice undergoes surface ablation. {\it (iii)} When the ice ablates entirely, the ocean mixed layer is represented as a thermodynamic reservoir using $T(t,E)=T_{ml}=E/(c_{ml}H_{ml})$. Using the ramp function as a convenient notation for combining cases {\it (i)} and {\it (ii)}, the surface temperature can be expressed as
\begin{equation}
T(t,E)= \begin{cases}
- \mathcal{R}\left[  \frac{ \left(1-\alpha_i \right) F_S(t) -F_0(t) + \Delta F_0 }{ k_i L_i/E - F_T(t) } \right] & E<0  \textrm{   [sea ice]}\\
\frac{E}{c_{ml}H_{ml}} & E \ge 0 \textrm{   [ocean]}
\end{cases} .
\label{eq:sum2}
\end{equation}

The ocean is represented as either ice-covered or ice-free at any given time. To model the gradual transition between these regimes in a partially ice-covered Arctic Ocean, the albedo varies between values for ice ($\alpha_i$) and ocean mixed layer ($\alpha_{ml}$) with a characteristic smoothness given by the thickness parameter $h_{\alpha}$,
\begin{equation}
  \alpha(E)= \frac{\alpha_{ml}+\alpha_i}{2} + \frac{\alpha_{ml}-\alpha_i}{2} \tanh\left(\frac{E}{L_i h_\alpha}\right).
\label{eq:sum3}
\end{equation}

We also consider a partially linearized version of the model in which equation \eqref{eq:sum2} is replaced with \begin{equation}
T(t,E)=\frac{E}{c_{ml}H_{ml}} 
\label{eq:linear}
\end{equation}
and there is no ice export ($v_0=0$). This causes the model equations to be linear with the exception of the ice-albedo feedback \eqref{eq:sum3}.

\section{Results}

\subsection{Seasonal cycle}

\begin{figure}
\centerline{\includegraphics[width=0.38\textwidth]{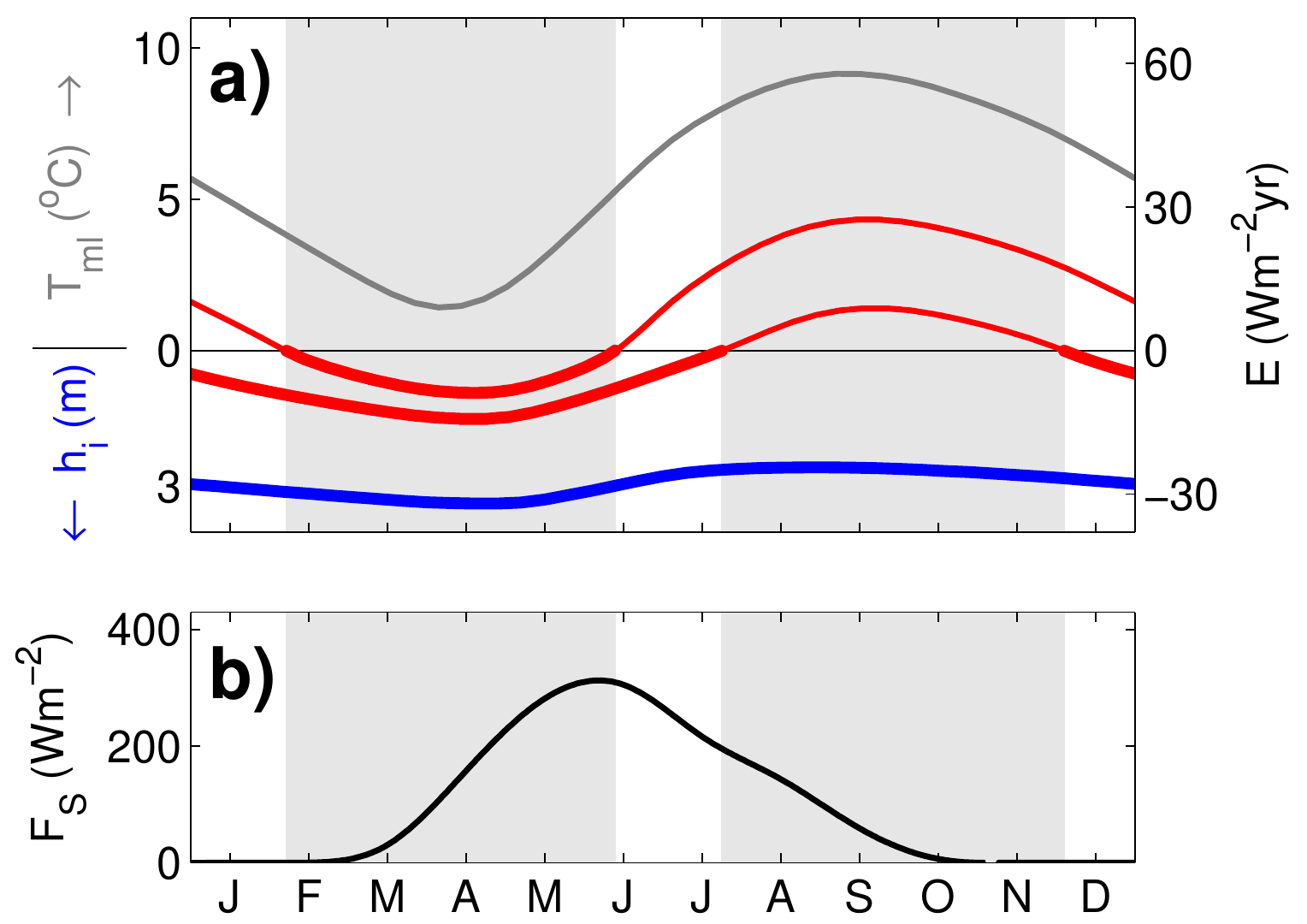}}
  \caption{Sea ice seasonal cycle in a warming climate and solar radiation. (a) Seasonal cycle of stable solutions of the full nonlinear model are illustrated by plotting the model state $E$ (energy per unit area in ocean mixed layer sensible heat or sea ice latent heat) versus time of year. Four solutions are plotted, each with different levels of surface heating $\Delta F_0$: a perennial ice state (blue curve, $\Delta F_0=0$), seasonally ice-free states with most of the year ice-covered (lower red curve, $\Delta F_0=21$ Wm$^{-2}$) or most of the year ice-free (upper red curve, $\Delta F_0=23$ Wm$^{-2}$), and a perennially ice-free state (gray curve, $\Delta F_0=19$ Wm$^{-2}$). As described in equation \eqref{eq:sum0}, when $E>0$, it represents the mixed layer temperature of an ice-free ocean ($E=c_{ml}H_{ml}T_{ml}$). At $E=0$, the ocean mixed layer reaches the freezing point ($T_{ml}=0$\mbox{$^\circ$}C), and further cooling will cause ice to grow. When $E<0$, it represents the sea ice thickness ($E=-L_i h_i$); note that ice thickness increases downward. Model solutions are drawn with thicker lines when the ocean is ice-covered and thinner lines when the ocean is ice-free.  Solutions are obtained by integrating equations \eqref{eq:sum1}-\eqref{eq:sum3} with seasonally varying parameter values given in Table S1 ({\it Supporting Information}) until the model has converged on a steady-state seasonal cycle. The light gray shaded region to the right represents the first months to become ice-free in a warming climate (demarcated by zero-crossings of the seasonally ice-free solution with $\Delta F_0=21$ Wm$^{-2}$), while the light gray shaded region to the left represents the last months that are ice-covered in a further warmed climate (demarcated by zero-crossings of the seasonally ice-free solution with $\Delta F_0=23$ Wm$^{-2}$). (b) Seasonal cycle of incident solar radiation specified in the model based on central Arctic surface observations \cite{Maykut-Untersteiner-1971:some}, indicating that the first months to become ice-free in a warming climate (light gray region to right) and the last months to be ice-covered in a further warmed climate (light gray region to left) experience similar amounts of solar radiation. Note that the radiation curve is asymmetric due to seasonal differences in Arctic cloudiness, but the qualitative results presented here do not depend on this asymmetry.}
\label{fig:sunlight-ice}
\end{figure}

In a seasonally varying Arctic climate, warming might be expected to cause the sea ice to initially melt back to the point where the entire Arctic Ocean is ice-free during part of the year, in contrast to the current perennial sea ice cover in the central Arctic. Further warming would cause the ice-free period to increase until the Arctic Ocean becomes perennially ice-free. We study this scenario theoretically by increasing the imposed surface heat flux $\Delta F_0$ in equations \eqref{eq:sum1}-\eqref{eq:sum3}. In Fig.~\ref{fig:sunlight-ice}a, steady-state seasonal cycle solutions are plotted in regimes with perennial ice cover (blue curve), seasonally ice-free conditions (red curves), and perennially ice-free conditions (gray curve).

The annual minimum sea ice area and thickness is commonly referred to as ``summer" sea ice and the annual maximum is commonly referred to as ``winter" sea ice. This nomenclature may carry with it the implication that the ice-albedo feedback, which depends on the magnitude of the incident solar radiation, would be most prominent during the retreat of the summer sea ice cover.  Indeed, it is often conjectured that a critical threshold for the loss of summer Arctic sea ice may be more likely than a threshold for the loss of winter ice \cite{Lenton-Held-Kriegler-et-al-2008:tipping}. 
However, as is illustrated by Fig.~\ref{fig:sunlight-ice}b, this terminology can be misleading because the ice cover receives a similar amount of incident solar radiation during the period of annual maximum as at annual minimum. The light gray shaded regions in Fig.~\ref{fig:sunlight-ice} illustrate the key transition periods in the state of the Arctic Ocean during the transition from perennial ice cover to seasonally ice-free conditions (gray region to right) and from seasonally ice-free conditions to perennially ice-free conditions (gray region to left). Both of these periods experience roughly equivalent amounts of incident solar radiation (Fig.~\ref{fig:sunlight-ice}b), with somewhat more solar radiation occurring during the period associated with the loss of winter ice (light gray region to left). Hence the ice-albedo feedback should be expected to be similarly strong during a transition to perennially ice-free conditions in a very warm climate (i.e., loss of winter ice) as during a more imminent possible warming to seasonally ice-free conditions (i.e., loss of summer ice).

\subsection{Bifurcation thresholds}

\begin{figure}[t]
\centerline{\includegraphics[height=0.30\textwidth]{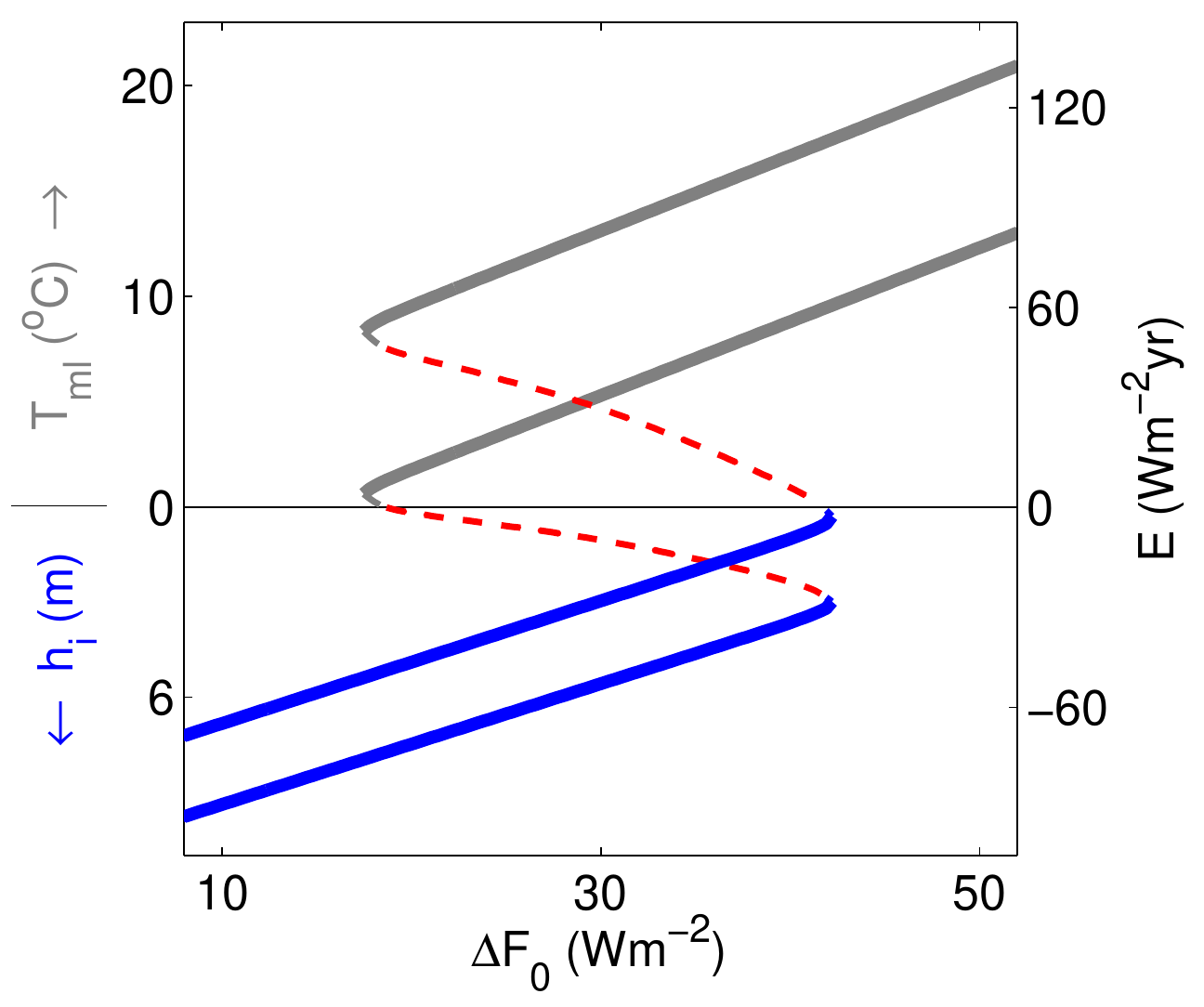}}
  \caption{Bifurcation diagram for the partially linearized model, where nonlinear sea ice thermodynamic effects have been excluded but the ice-albedo feedback has been retained (equations \eqref{eq:sum1}-\eqref{eq:sum2}, \eqref{eq:linear}). For each value of the surface heating $\Delta F_0$, the model is integrated until it converges on a steady-state seasonal cycle, and the annual maximum (upper curve) and annual minimum (lower curve) values of $E$ are plotted. Solutions with perennial sea ice cover are indicated in blue, seasonally ice-free solutions in red, and perennially ice-free solutions in gray. Dashed lines indicate unstable solutions, which have been located by constructing an annual Poincar{\'e} map and finding the fixed points (i.e., numerically integrating the model for one year starting from an array of initial conditions and identifying the solutions with the same value of $E$ at the end of the year as the initial condition). The curves have been smoothed with a boxcar filter to suppress a small level of noise associated with numerical integration. Note that the lines are slightly curved at the two bifurcation points due to the smooth albedo transition associated with $h_\alpha>0$. The vertical axis is labeled as in Fig.~\ref{fig:sunlight-ice}a.}
\label{fig:bifn1}
\end{figure}

We begin the bifurcation analysis using the partially linearized version of the model (equations \eqref{eq:sum1}, \eqref{eq:sum3}-\eqref{eq:linear}) to focus on the effect of albedo in the absence of other nonlinearities. In this representation, the Arctic Ocean is viewed as a simple radiating thermal reservoir with a temperature-dependent albedo, and the model exhibits a linear relaxation to a stable solution in each albedo regime. As would be expected by analogy with the discussion above of an annual mean Arctic Ocean with a variable sea ice edge, Fig.~\ref{fig:bifn1} illustrates that when $\Delta F_0$ becomes sufficiently large for the ocean to remain perennially ice-free with $\alpha=\alpha_{ml}$, an unstable seasonally ice-free solution (red dashed curve) appears in a saddle-node bifurcation of cycles (for a discussion of the theory of bifurcations in periodic systems, see, e.g., Strogatz \cite{Strogatz-1994:nonlinear}).
The unstable solution separates stable solutions with perennial ice (blue curve) or perennially ice-free conditions (gray curve). The perennial ice regime collides with the unstable seasonally ice-free state and disappears in a second saddle-node bifurcation of cycles at the point where $\Delta F_0$ becomes sufficiently large that the ice completely melts at the time of annual maximum $E$ in the cold stable state. Due to there being significant incident solar radiation during both the maximum and minimum periods of the seasonal cycle of $E$ (Fig.~\ref{fig:sunlight-ice}), the ice-albedo feedback ensures that all seasonally ice-free solutions will be unstable (Fig.~\ref{fig:bifn1}).

When nonlinear sea ice thermodynamic effects are included (equations \eqref{eq:sum1}-\eqref{eq:sum3}), basal ice formation is controlled by a diffusive vertical heat flux of $k_i\Delta T/h_i$, where $\Delta T$ is the difference between surface and basal temperatures and the base is assumed to be at the freezing point. This causes thin ice to grow significantly faster than thick ice \cite{Maykut-Untersteiner-1971:some}. It would also cause thin ice to experience greater basal ablation during the summer melt season, but the surface temperature only warms until it reaches the freezing point ($\Delta T=0$) and surface melt begins, making the rate of melt less sensitive to thickness. These two effects, both nonlinear in $E$, are expressed in equation \eqref{eq:sum2} by the $-k_i/h=k_i L_i/E$ term in the denominator and the ramp function $\mathcal{R}(x)$, respectively. The result is an increase in the rate of growth for thin ice which is more stabilizing for thinner ice, as pointed out \cite{Maykut-1986:surface} and applied \cite{Bitz-Roe-2004:mechanism} in previous studies. This is in contrast to the state-independent linear mixed layer stabilizing term, $-F_T(t) E/c_{ml}H_{ml}$, which applies when $E>0$ (equations \eqref{eq:sum1} and \eqref{eq:sum2}).

\begin{figure}
\centerline{\includegraphics[height=0.30\textwidth]{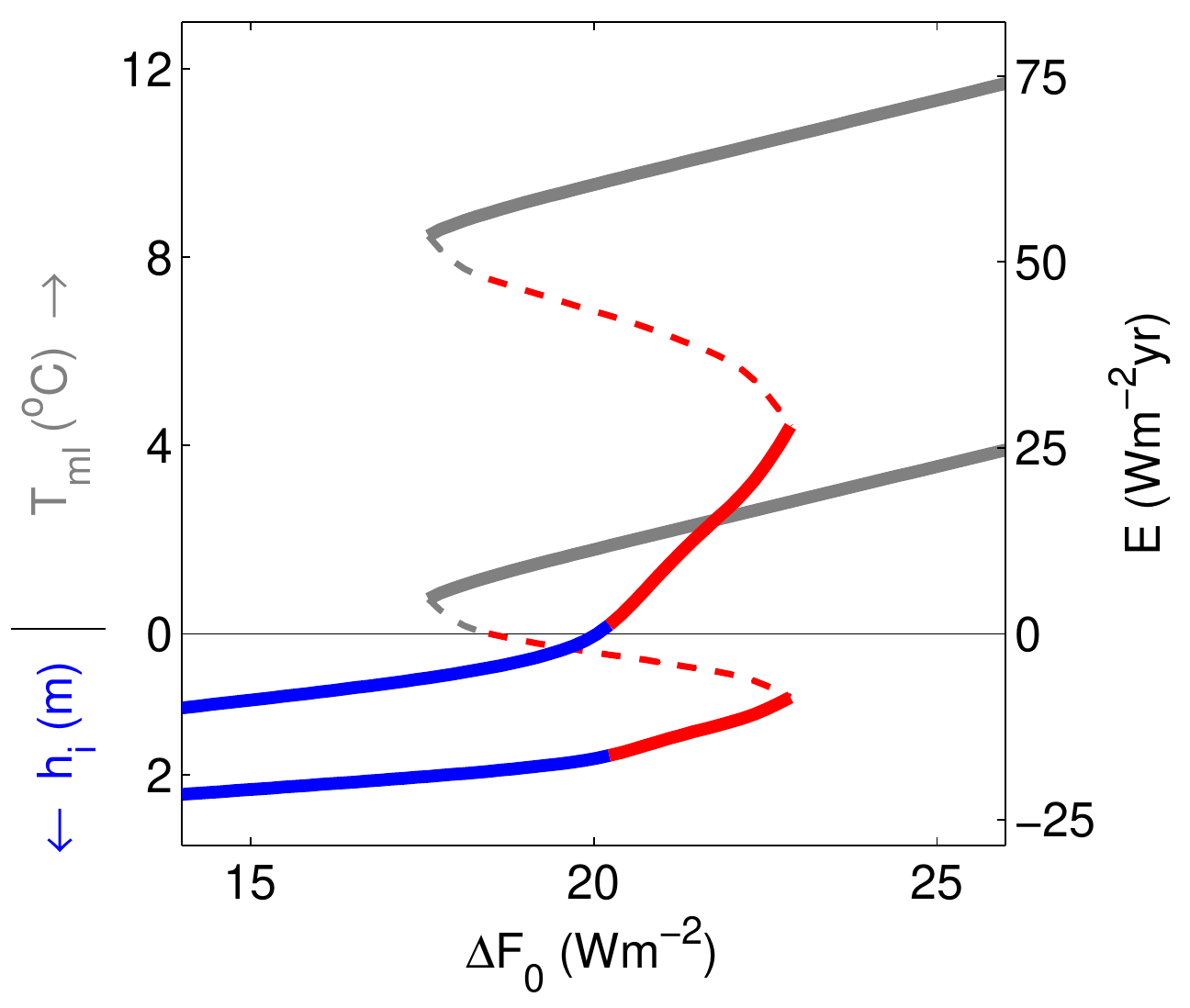}}
  \caption{Bifurcation diagram for the full nonlinear model (equations \eqref{eq:sum1}-\eqref{eq:sum3}). Axes and colors are as described in the caption of Fig.~\ref{fig:bifn1}. The inclusion of nonlinear sea ice thermodynamic effects stabilizes the model when sea ice is present during a sufficiently large fraction of the year, allowing stable seasonally ice-free solutions (red solid curves). Under a modest warming ($\Delta F_0=15$ Wm$^{-2}$), modeled sea ice thickness varies seasonally between 0.9m and 2.2m. Further warming ($\Delta F_0=20$ Wm$^{-2}$) causes the September ice cover to disappear, and the system undergoes a smooth transition to seasonally ice-free conditions. When the model is further warmed ($\Delta F_0=23$ Wm$^{-2}$), a saddle-node bifurcation occurs, and the wintertime sea ice-cover abruptly disappears in an irreversible process. While the specific values of $\Delta F_0$ at which the transitions occur are sensitive to parameter choices, the qualitative features of Fig.~\ref{fig:bifn2} are highly robust to changes in model parameter values ({\it Supporting Information} Fig.~S4).}
\label{fig:bifn2}
\end{figure}

These nonlinearities allow for the existence of a stable seasonally ice-free solution (Fig.~\ref{fig:bifn2}). When a sufficiently large value of $\Delta F_0$ is chosen such that the cold solution becomes ice-free during a small part of the year, a slight increase in temperature would lead to a longer open-water period and a thinner seasonal ice cover. Although the increased period of open water promotes warming through the ice-albedo feedback, the thinner ice grows significantly faster because of the sea ice thermodynamic effects which are nonlinear in $E$. During the ice-covered portion of the year, the stability of the solution is controlled by this strong nonlinear stabilizing effect, but during the ice-free portion of the year it is replaced by the weaker linear mixed layer stabilizing term. This causes the stabilizing sea ice thermodynamic effects to dominate the destabilizing ice-albedo feedback and allow a stable seasonally ice-free solution only when there is ice cover during a sufficiently long portion of the year. Nonetheless, the ice-albedo feedback causes this regime to warm at an increased rate in response to increasing heat flux (compare slopes of red and blue curves in Fig.~\ref{fig:bifn2}). As the ice-covered fraction of the year decreases in a warming climate, the stabilizing ice thermodynamic effects become less pronounced in the full annual cycle, and a bifurcation occurs when ice covers the Arctic Ocean during a sufficiently small fraction of the year to allow the ice-albedo feedback to dominate. Hence when the Arctic warms beyond this point, the system supports only an ice-free solution (Fig.~\ref{fig:bifn2}).

\section{Discussion}

\subsection{Comparison with results of other models}

The theoretical treatment presented here is constructed to facilitate simple conceptual interpretation, and to this end many processes have been neglected. Factors including possible sea ice--cloud feedbacks \cite{Kellogg-1975:climatic, 
Climate-2003:understanding, Vavrus-2004:impact,
Abbot-Tziperman-2008:winter}, the dependence of sea ice surface albedo on snow and melt pond coverage \cite{Curry-Schramm-Ebert-1995:sea, Flato-Brown-1996:variability},
ocean heat flux convergence feedbacks \cite{Holland-Bitz-Tremblay-2006:future, Winton-2006:does}, changes in wind-driven ice dynamics \cite{Maslanik-Stroeve-Drobot-et-al-2007:younger}, and changes in ice rheology \cite{Zhang-Rothrock-2005:effect} in a thinning ice cover \cite{Rothrock-Percival-Wensnahan-2008:decline} could potentially lead to other bifurcation thresholds or smooth out the threshold investigated here,  akin to the smoothing of a first order phase transition due to statistical fluctuations \cite{StatPhys}. We are emboldened in our approach, however, because behavior consistent with the mechanism proposed here can be found in the previously published results of models with a broad range of complexities. {\it (i)} A ``toy model'' which is forced by a step function seasonal cycle produced no stable seasonally ice-free solution in the published parameter regime \cite{Thorndike-1992:toy}, but by a slight adjustment of the tunable model parameters one can find a stable seasonally ice-free solution which coexists with a stable perennially ice-free solution ({\it Supporting Information} Fig.~S5), consistent with the findings presented here. {\it (ii)} In a variant of the model used in this study that is significantly more complex (representing the simultaneous evolution of fractional Arctic sea ice coverage, mean thickness, and surface temperature, as well as ocean mixed layer temperature), increasing the level of greenhouse gas forcing leads to a gradual transition to seasonally ice-free solutions followed by a bifurcation threshold during the transition to perennially ice-free conditions \cite{Eisenman-2007:arctic}, as in Fig.~\ref{fig:bifn2}. {\it (iii)} Turning to the most complex current climate models, about half the coupled atmosphere--ocean global climate models used for the most recent IPCC report \cite{Ipcc-2007:climate} predict seasonally ice-free Arctic Ocean conditions by the end of the 21st century, and none predict perennially ice-free conditions by the end of the 21st century. However, perennially ice-free Arctic Ocean conditions occur in two of the model simulations after CO$_2$ quadrupling. Neither of the models exhibits an abrupt transition when the annual minimum (September) ice cover disappears, but after further warming one of the models abruptly loses its March ice cover when it becomes perennially ice-free \cite{Winton-2006:does}. The physical mechanism presented here may help explain this abrupt loss of simulated March ice while the simulated September ice receded gradually.

\subsection{Conclusions}

Our analysis suggests that a sea ice bifurcation threshold (or ``tipping point'') caused by the ice-albedo feedback is not expected to occur in the transition from current perennial sea ice conditions to a seasonally ice-free Arctic Ocean, but that a bifurcation threshold associated with the sudden loss of the remaining seasonal ice cover may occur in response to further heating. These results may be interpreted by viewing the state of the Arctic Ocean as comprising a full seasonal cycle, which can include ice-covered periods as well as ice-free periods. The ice-albedo feedback promotes the existence of multiple states, allowing the possibility of abrupt transitions in the sea ice cover as the Arctic is gradually forced to warm. Because a similar amount of solar radiation is incident at the surface during the first months to become ice-free in a warming climate as during the final months to lose their ice in a further warmed climate, the ice-albedo feedback is similarly strong during both transitions. The asymmetry between these two transitions is associated with the fundamental nonlinearities of sea ice thermodynamic effects, which make the Arctic climate more stable when sea ice is present than when the open ocean is exposed.  Hence when sea ice covers the Arctic Ocean during fewer months of the year, the state of the Arctic becomes less stable and more susceptible to destabilization by the ice-albedo feedback. In a warming climate, as discussed above, this causes irreversible threshold behavior during the potential distant loss of winter ice, but not during the more imminent possible loss of summer (September) ice.

The relevance of any basic theory to the actual future evolution of the complex climate system must be carefully qualified. Since the time scale associated with the sea ice response to a change in forcing may be decadal, and the time scale associated with increasing greenhouse gas concentrations may be similar, the system may not be operating close to a steady-state. In the gradual approach to steady-state under a continual change in forcing, the difference between a region of the steady-state solution with increased sensitivity to the forcing and an actual discontinuous bifurcation threshold (as in Fig.~\ref{fig:bifn2}) could be difficult to discern. If greenhouse gas concentrations were reduced after crossing a bifurcation threshold, however, the possible irreversibility of the trajectory would certainly be expected to be relevant.



\begin{acknowledgments}
The authors are grateful to the Geophysical Fluid Dynamics summer program at Woods Hole Oceanographic Institution (NSF OCE0325296) where the development of the physical representations employed in this study benefited from discussions with many visitors and staff including Norbert Untersteiner, John Walsh, Jamie Morison, Dick Moritz, Danny Feltham, G\"oran Bj\"ork, Bert Rudels, Doug Martinson, Andrew Fowler, George Veronis, Grae Worster, Neil Balmforth,  Ed Spiegel, Joe Keller, and Alan Thorndike. IE thanks Eli Tziperman and Cecilia Bitz for helpful conversations during the course of this work. The authors thank Richard Goody, Tapio Schneider, and Eli Tziperman for comments on the manuscript. JSW acknowledges support from NSF OPP0440841 and Yale University, USA, and the Wenner-Gren Foundation, the Royal Institute of Technology, and NORDITA in Stockholm. IE acknowledges support from a NASA Earth and Space Science Fellowship, a WHOI Geophysical Fluid Dynamics Fellowship, NSF paleoclimate program grant ATM-0502482, the McDonnell Foundation, a prize postdoctoral fellowship through the California Institute of Technology Division of Geological and Planetary Sciences, and a NOAA Climate and Global Change Postdoctoral Fellowship administered by the University Corporation for Atmospheric Research.
\end{acknowledgments}

\bibliographystyle{pnas}

\begin{thebibliography}{10}

\bibitem{Stroeve-Serreze-Fetterer-et-al-2005:tracking}
Stroeve, JC et~al.
\newblock (2005) Tracking the {Arctic's} shrinking ice cover: Another extreme
  {September} minimum in 2004.
\newblock {\em Geophys Res Lett} 32:L04501.

\bibitem{Perovich-Light-Eicken-et-al-2007:increasing}
Perovich, DK et~al.
\newblock (2007) Increasing solar heating of the {A}rctic {O}cean and adjacent
  seas, 1979-2005: Attribution and role in the ice-albedo feedback.
\newblock {\em Geophys Res Lett} 34:L19505.

\bibitem{Lindsay-Zhang-2005:thinning}
Lindsay, RW, Zhang, J
\newblock (2005) The thinning of {Arctic} sea ice, 1988-2003: Have we passed a
  tipping point?
\newblock {\em J Clim} 18:4879--4894.

\bibitem{Overpeck-Sturm-Francis-et-al-2005:arctic}
Overpeck, J et~al.
\newblock (2005) Arctic system on trajectory to new, seasonally ice-free state.
\newblock {\em EOS} 86:309--313.

\bibitem{Serreze-Francis-2006:arctic}
Serreze, MC, Francis, JA
\newblock (2006) The {Arctic} amplification debate.
\newblock {\em Clim Change} 76:241--264.

\bibitem{Holland-Bitz-Tremblay-2006:future}
Holland, MM, Bitz, CM, Tremblay, B
\newblock (2006) Future abrupt reductions in the summer {A}rctic sea ice.
\newblock {\em Geophys Res Lett} 33:L23503.

\bibitem{Maslanik-Stroeve-Drobot-et-al-2007:younger}
Maslanik, J et~al.
\newblock (2007) A younger, thinner arctic ice cover: Increased potential for
  rapid, extensive sea-ice loss.
\newblock {\em Geophys Res Lett} 34:L24501.

\bibitem{Lenton-Held-Kriegler-et-al-2008:tipping}
Lenton, TM et~al.
\newblock (2008) {Tipping elements in the Earth's climate system}.
\newblock {\em Proc Nat Acad Sci USA} 105:1786--1793.

\bibitem{Merryfield-Holland-Monahan-2006:multiple}
Merryfield, W, Holland, M, Monahan, A
\newblock (2008) in {\em Arctic Sea Ice Decline: Observations, Projections,
  Mechanisms, and Implications}, eds{} Bitz, C, {DeWeaver}, E
\newblock (Am Geophys Union), in press.

\bibitem{Budyko-1969:effect}
Budyko, MI
\newblock (1969) The effect of solar radiation variations on the climate of the
  earth.
\newblock {\em Tellus} 21:611--619.

\bibitem{Sellers-1969:global}
Sellers, WD
\newblock (1969) A global climate model based on the energy balance of the
  earth-atmosphere system.
\newblock {\em J Appl Meteor} 8:392--400.

\bibitem{North-1990:multiple}
North, GR
\newblock (1990) Multiple solutions in energy-balance climate models.
\newblock {\em Global Planet Change} 82:225--235.

\bibitem{Maykut-Untersteiner-1971:some}
Maykut, GA, Untersteiner, N
\newblock (1971) Some results from a time-dependent thermodynamic model of sea
  ice.
\newblock {\em J Geophys Res} 76:1550--1575.

\bibitem{Maykut-Church-1973:radiation}
Maykut, GA, Church, PE
\newblock (1973) Radiation climate of {Barrow}, {Alaska}, 1962-66.
\newblock {\em J Appl Meteor} 12:620--628.

\bibitem{Kalnay-Kanamitsu-Kistler-et-al-1996:ncep}
Kalnay, E et~al.
\newblock (1996) The {NCEP/NCAR} 40-year reanalysis project.
\newblock {\em Bull Amer Meteor Soc} 77:437--471.

\bibitem{Nakamura-Oort-1988:atmospheric}
Nakamura, N, Oort, AH
\newblock (1988) Atmospheric heat budgets of the polar-regions.
\newblock {\em J Geophys Res} 93:9510--9524.

\bibitem{Kwok-Cunningham-Pang-2004:fram}
Kwok, R, Cunningham, GF, Pang, SS
\newblock (2004) Fram strait sea ice outflow.
\newblock {\em J Geophys Res} 109:C01009.

\bibitem{Strogatz-1994:nonlinear}
Strogatz, SH
\newblock (1994) {\em Nonlinear Dynamics and Chaos}
\newblock (Perseus Books).

\bibitem{Maykut-1986:surface}
Maykut, G
\newblock (1986) in {\em The geophysics of sea ice}, ed{} Untersteiner, N
\newblock (Plenium), pp 395--463.

\bibitem{Bitz-Roe-2004:mechanism}
Bitz, CM, Roe, GH
\newblock (2004) A mechanism for the high rate of sea ice thinning in the
  arctic ocean.
\newblock {\em J Clim} 17:3623--3632.

\bibitem{Kellogg-1975:climatic}
Kellogg, WW
\newblock (1975) in {\em Climate of the Arctic}, eds{} Weller, G, Bowling, SA
\newblock (Geophys Inst, Univ Alaska Fairbanks), pp 111--116.

\bibitem{Climate-2003:understanding}
{Panel on Climate Change Feedbacks}
\newblock (2003) {\em Understanding Climate Change Feedbacks}
\newblock (National Academies Press).

\bibitem{Vavrus-2004:impact}
Vavrus, S
\newblock (2004) The impact of cloud feedbacks on arctic climate under
  greenhouse forcing.
\newblock {\em J Clim} 17:603--615.

\bibitem{Abbot-Tziperman-2008:winter}
Abbot, DS, Tziperman, E
\newblock (2008) Winter {A}rctic sea ice uncertainty under global warming due
  to a cloud radiative feedback.
\newblock {Submitted to {\em J Clim}}.

\bibitem{Curry-Schramm-Ebert-1995:sea}
Curry, JA, Schramm, JL, Ebert, EE
\newblock (1995) Sea-ice albedo climate feedback mechanism.
\newblock {\em J Clim} 8:240--247.

\bibitem{Flato-Brown-1996:variability}
Flato, GM, Brown, RD
\newblock (1996) Variability and climate sensitivity of landfast {Arctic} sea
  ice.
\newblock {\em J Geophys Res} 101:25767--25777.

\bibitem{Winton-2006:does}
Winton, M
\newblock (2006) Does the arctic sea ice have a tipping point?
\newblock {\em Geophys Res Lett} 33:L23504.

\bibitem{Zhang-Rothrock-2005:effect}
Zhang, JL, Rothrock, DA
\newblock (2005) Effect of sea ice rheology in numerical investigations of
  climate.
\newblock {\em J Geophys Res} 110: C08014.

\bibitem{Rothrock-Percival-Wensnahan-2008:decline}
Rothrock, DA, Percival, DB, Wensnahan, M
\newblock (2008) The decline in {A}rctic sea-ice thickness: Separating the
  spatial, annual, and interannual variability in a quarter century of
  submarine data.
\newblock {\em J Geophys Res} 113:C05003.

\bibitem{Thorndike-1992:toy}
Thorndike, AS
\newblock (1992) A toy model linking atmospheric thermal radiation and sea ice
  growth.
\newblock {\em J Geophys Res} 97:9401--9410.

\bibitem{Eisenman-2007:arctic}
Eisenman, I
\newblock (2007) in {\em 2006 Program of Studies: Ice (Geophysical Fluid
  Dynamics Program)}
\newblock (Woods Hole Oceanog. Inst. Tech. Rept. 2007-02), pp 133--161
\newblock http://gfd.whoi.edu/page.do?pid=12938.

\bibitem{Ipcc-2007:climate}
Solomon, S et~al., eds
\newblock (2007) {\em {Climate change 2007: The Physical Science Basis.
  Contribution of Working Group I to the Fourth Assessment Report of the
  Intergovernmental Panel on Climate Change}}
\newblock (Cambridge University Press), p 996.

\bibitem{StatPhys}
Lifshitz, EM, Pitaevskii, LP 
\newblock (1980) Statistical physics 
\newblock(Pergamon, Oxford).

\end{thebibliography}

\end{article}

\end{document}


\title{SI Appendix}

\author{Eisenman and Wettlaufer 10.1073/pnas.0806887106\affil{1}{~}}

\contributor{}

\maketitle

\begin{article}

\renewcommand{\deg}{\mbox{$^\circ$}}
\newcommand {\Wm}{Wm$^{-2}$}
\newcommand {\lt}{\left(}
\newcommand {\rt}{\right)}

\setcounter{equation}{5}
\renewcommand{\thefigure}{S\arabic{figure}}




Here we derive the idealized Arctic sea ice--ocean--atmosphere model that is summarized in equations {\bf [1]}-{\bf [4]} of the Research Report. Note that we carry out the entire derivation using dimensional variables, rather than following the conventional mathematical development of such equations through a process of non-dimensionalization, in order to make direct contact with previous studies of the thermodynamics of sea ice and climate.

\section{Sea Ice}

The evolution of the sea ice temperature profile is an idealized version of the single-column thermodynamic model of Maykut and Untersteiner \cite{Maykut-Untersteiner-1971:some} (hereafter MU71). Vertical heat conduction in sea ice is computed in MU71 according to
\begin{equation}
  c_{\textrm{eff}}(\tilde{T},S) \frac{\partial \tilde{T}}{\partial t} = 
  \frac{\partial}{\partial z} \left[ k_{\textrm{eff}}(\tilde{T},S)\frac{\partial \tilde{T}}{\partial z} \right]
   +A_R ,
  \label{eq:MU71-T-si}
\end{equation}
which can be derived from the general theory of mushy layers \cite{Feltham-Untersteiner-Wettlaufer-et-al-2006:sea}. Here $A_R$ represents the absorption of shortwave radiation that has penetrated below the surface of the ice, the effective heat capacity $c_{\textrm{eff}}(\tilde{T},S)$ and thermal conductivity $k_{\textrm{eff}}(\tilde{T},S)$ depend on simulated temperature $\tilde{T}$ and specified salinity $S$, and the vertical coordinate $z$ increases upward. Note that for the $\tilde{T}$ and $S$ range in perennial ice, MU71 neglect the vertical derivative of the effective conductivity, $\partial k_{\textrm{eff}}(\tilde{T},S) / \partial z$, allowing the first term on the right-hand side of equation \eqref{eq:MU71-T-si} to be expressed as $k_{\textrm{eff}}(\tilde{T},S) \partial^2 \tilde{T} / \partial z^2$. MU71 also include a layer of snow above the ice with specified snowfall and simulated snow melt.

The boundary condition in MU71 at the upper surface ($z=h_T$) is a flux balance when the ice is below the freezing temperature ($\tilde{T}_{fr}$) and otherwise a Stefan condition for surface ablation:
\begin{equation}
  \left[k_{\textrm{eff}}(\tilde{T},S)\frac{\partial \tilde{T}}{\partial z}\right]_{z={h_T}}+F_{top}(t,T_i,\alpha_i)
  =\begin{cases}0 & T_i < 0 \\
   L_{i}\frac{d h_T}{dt} & T_i = 0 \end{cases} ,
\label{eq:MU71top}
\end{equation}
with $L_i$ the latent heat of fusion of ice, $\alpha_i$ the surface albedo, $T_i \equiv \tilde{T}_i-\tilde{T}_{fr}$ the surface temperature departure from the freezing point with $\tilde{T}_i\equiv \tilde{T}(z=h_T)$, and $F_{top}(t,T_i,\alpha_i)$ representing the sum of sensible and latent heat fluxes and longwave and shortwave radiative fluxes out of the surface. The seasonal cycle of each of these components of the surface flux are specified in MU71 based on observations, except for the upward longwave flux which is computed from the surface temperature using the Stefan-Boltzmann equation. To facilitate an analytical solution for $T_i$ (equation \eqref{eq:ice0} below), we approximate the Stefan-Boltzmann equation by its linearized version, $\sigma \tilde{T}_i^4 =  \sigma_0+\sigma_T T_i $, where $\sigma$ is the Stefan-Boltzmann constant and the parameters ($\sigma_0=316$ Wm$^{-2}$, $\sigma_T=3.9$ Wm$^{-2}$K$^{-1}$) are chosen such that the equation is exact when $\tilde{T}_i=-30$\deg C and when $\tilde{T}_i=0$\deg C, which are the approximate values of $\tilde{T}_i$ during most of the winter and summer, respectively. This allows the temperature dependence of the surface flux to be expressed as 
\begin{equation}
F_{top}(t,T_i,\alpha_i)= -(1-\alpha_i) F_S(t) + F_0(t) + F_T(t) T_i ,
\label{eq:Ftop}
\end{equation}
where $F_S(t)$ is the downwelling shortwave radiation flux, $F_0(t)$ is $\sigma_0$ plus the specified sensible and latent heat fluxes, and $F_T(t)=\sigma_T$. Note that here the atmosphere is specified as in MU71, whereas in the full coupled version of the model $F_0(t)$ and $F_T(t)$ take on a different set of values computed using the atmospheric model (equations \eqref{eq:atm05}-\eqref{eq:atm06} below).

At the ice--ocean interface ($z=h_B$), MU71 apply a Stefan condition for ice growth or ablation,
\begin{equation}
  -L_i\frac{d h_B}{d t}  =
   -\left[ k_{\textrm{eff}}(\tilde{T},S) \frac{\partial \tilde{T}}{\partial z} \right]_{z={h_B}}
    -F_B,
\label{eq:MU71bot}
\end{equation}
with the flux from the ocean into the base of the ice specified to take a constant value of $F_B=2$ \Wm. Note that the temperature at the ice--ocean interface must be at the freezing point, $\tilde{T}(z={h_B})=\tilde{T}_{fr}$. The upper and lower surfaces of the ice, $h_T$ and $h_B$, evolve separately in MU71, who use a coordinate system in which each ice parcel remains stationary, and the predicted ice thickness is $h_i=h_{T}-h_B$ (see schematic in Fig.\  S1).

Here we neglect snow (MU71 report that having no snow causes the annual mean thickness to increase by 17cm from the standard case value of 288cm), and we neglect penetrating shortwave radiation $A_R\to 0$ (which MU71 report causes the annual mean ice thickness to decrease by 45cm). The impact of neglecting both of these factors is shown in Fig.\ S2 (black curves). 

\begin{figure}[b]
\centerline{\includegraphics[width=0.25\textwidth]{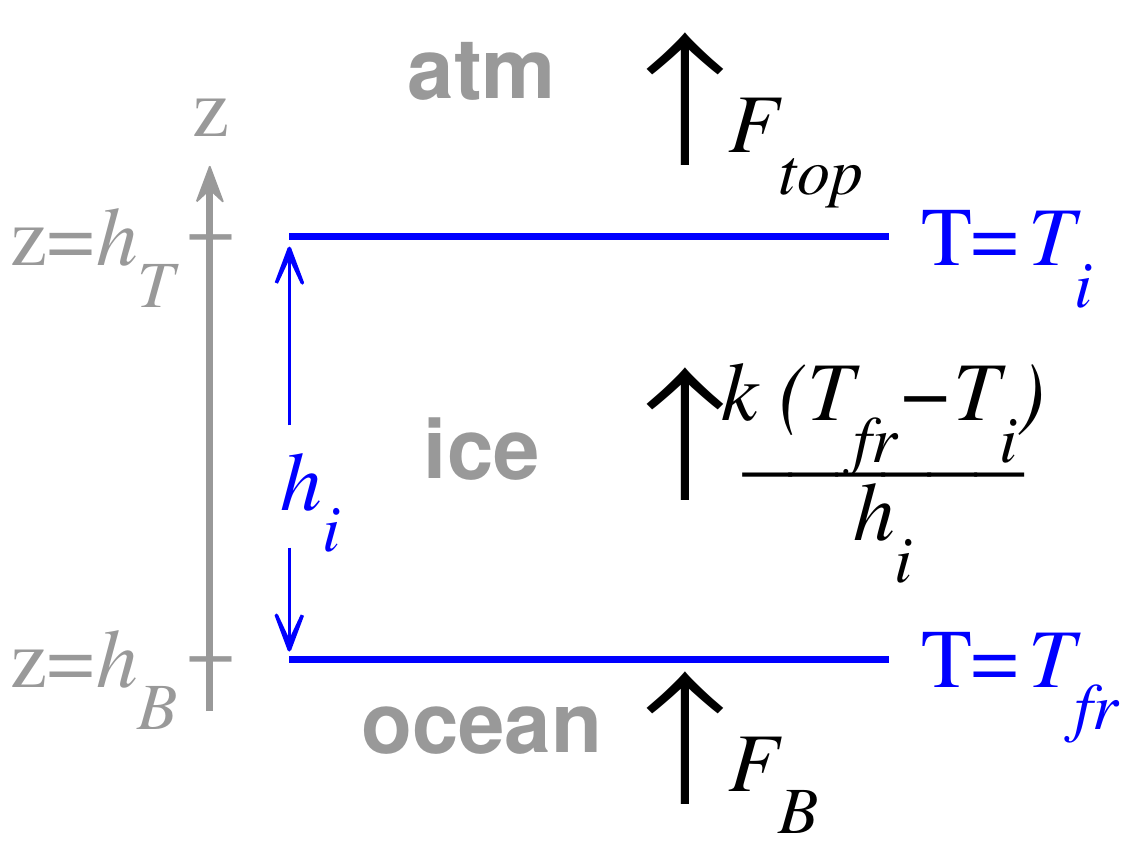}}
\caption{Schematic showing fluxes and variables in the sea ice component of the idealized model presented here. All fluxes are defined such that a positive value implies an upward flux.}
\label{fig:MU71-fluxes}
\end{figure}

\begin{figure*}
\centerline{\includegraphics[width=0.85\textwidth]{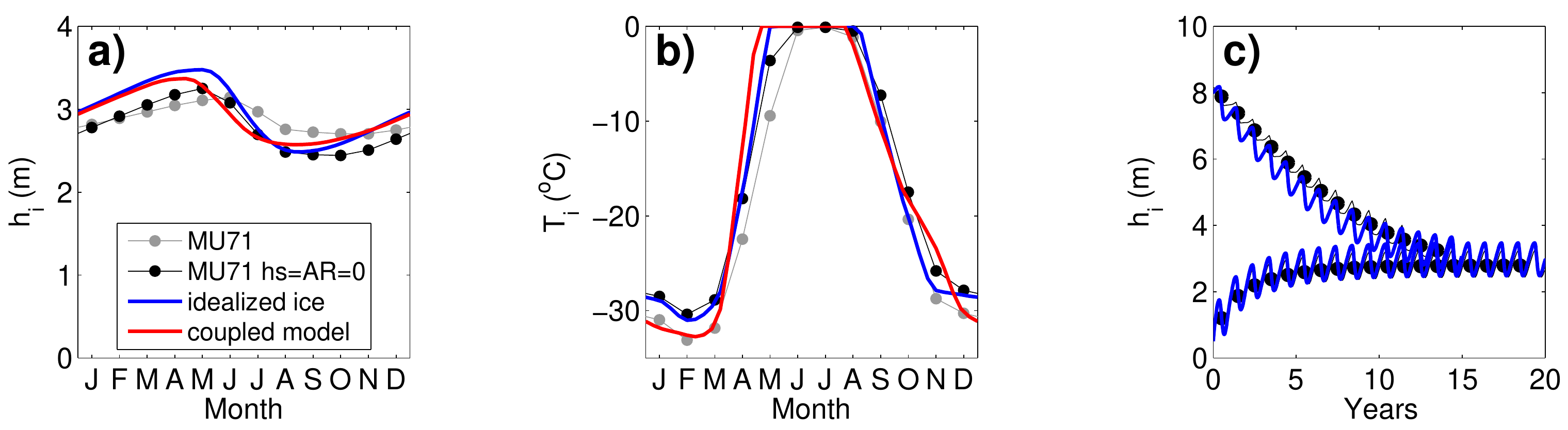}}
\caption{Effects of approximating the ice thermodynamics in the model of MU71  \cite{Maykut-Untersteiner-1971:some}. (a) Steady-state solution seasonal cycle of ice thickness in the MU71 standard case simulation (gray curve and circles), in a simulation with the MU71 model carried out for this study with no snow or penetrating shortwave radiation ($h_s=A_R=0$; black curve and circles), when the MU71 representation is replaced by the idealized sea ice model given by equations \eqref{eq:ice0}-\eqref{eq:ice1} (blue curve), and the standard case run with the fully coupled idealized sea ice--ocean--atmosphere model summarized in equations  {\bf [1]}-{\bf [4]} of the Research Report (red curve). (b) Seasonal cycle of surface temperature for the same four solutions as in (a).
Note that the surface temperature in the idealized model is diagnosed from the computed thickness and the specified surface forcing.
(c) Relaxation time to reach steady-state ice thickness from two different initial conditions in the MU71 model with $h_s=A_R=0$ (black curves and circles) and in the idealized sea ice model (blue curves).}
\label{fig:idealized-ice}
\end{figure*}

The thermal conductivity $k_{\textrm{eff}}(\tilde{T},S)$ in the MU71 standard case run is always 90\%--100\% of the pure ice value, and we approximate it to take the constant pure ice value,  $k_{\textrm{eff}}(\tilde{T},S)=k_i=2$ Wm$^{-1}$K$^{-1}$. The freezing temperature in MU71 is taken to be $\tilde{T}_{fr}=-1.8$\deg C at the base of the ice and $\tilde{T}_{fr}=-0.1$\deg C at the upper ice surface, and we approximate it to take a constant value of $\tilde{T}_{fr}=0$\deg C. Lastly, because the Stefan number $N_S\equiv L_i/\left[ c_{\textrm{eff}}(\tilde{T},S) \Delta \tilde{T} \right]$ is large, the temperature field in the ice relaxes quickly in response to changes in the solidification rate.  The actual values of $N_S$ predicted by the MU71 standard case seasonal cycle vary widely but typically are $N_S \gg 1$. Under these conditions, the system \eqref{eq:MU71-T-si}-\eqref{eq:MU71top}, \eqref{eq:MU71bot} can be expressed as a single ordinary differential equation, as described below.

Applying the large Stefan number approximation to the heat conduction equation \eqref{eq:MU71-T-si} yields a linear temperature profile, $\tilde{T} =  \tilde{T}_{fr} + (\tilde{T_i}-\tilde{T}_{fr})(z-h_B)/(h_T-h_B) = \tilde{T}_{fr} + T_i (z-h_B)/h_i$, which can be derived using a scaling argument after vertically integrating equation \eqref{eq:MU71-T-si} and inserting the boundary conditions \eqref{eq:MU71top}, \eqref{eq:MU71bot}. Next, this quasi-stationary temperature field is inserted into equations \eqref{eq:MU71top}, \eqref{eq:MU71bot}. The upper boundary condition \eqref{eq:MU71top} includes two different cases depending on whether or not surface ablation is occurring:

{\it (i)} When the surface is below the freezing temperature ($T_i<0$), the upper boundary condition \eqref{eq:MU71top} with the linear temperature profile takes the form
\begin{equation}
k_i \frac{T_i}{h_i} =-F_{top}(t,T_i,\alpha_i) .
\label{eq:T92-freeze1a}
\end{equation}
Since there is no surface ablation ($dh_T/dt=0$), the ice thickness evolves as $d h_i/d t=d/dt\lt h_T-h_B \rt=-dh_B/dt$, and after inserting the linear temperature profile the lower boundary condition \eqref{eq:MU71bot} becomes
  \begin{equation}
   L_i\frac{d h_i}{dt}=-k_i\frac{T_i}{h_i}-F_B .
  \label{eq:T92-freeze2}
\end{equation}
Inserting the surface temperature \eqref{eq:T92-freeze1a} into equation \eqref{eq:T92-freeze2} shows that ice thickness evolves according to
\begin{equation}
  L_i\frac{d h_i}{dt}=F_{top}(t,T_i,\alpha_i)-F_B .
\label{eq:T92-freeze2b}
\end{equation}

{\it (ii)} During surface ablation ($T_i= 0$), the temperature profile takes on a constant value of $\tilde{T}=\tilde{T}_{fr}$. Hence the upper boundary condition \eqref{eq:MU71top} takes the form
\begin{equation}
F_{top}(t,T_i,\alpha_i)=L_i \frac{dh_T}{dt} ,
\label{eq:T92-melt1}
\end{equation}
and the lower boundary condition \eqref{eq:MU71bot} becomes
\begin{equation}
-L_i \frac{dh_B}{dt}=-F_B ,
\label{eq:T92-melt2}
\end{equation}
which together imply that ice thickness, $h_i=h_{T}-h_B$, evolves according to an equation identical to the case with the surface below the freezing temperature \eqref{eq:T92-freeze2b}.

The steady-state surface temperature can be derived by inserting equation \eqref{eq:Ftop} into equation \eqref{eq:T92-freeze1a} to yield an algebraic solution for the case with $T_i<0$, which can be combined with the ablation case ($T_i=0$) as
\begin{equation}
T_i(t,h_i)= - \mathcal{R}\left[ \frac{ \left( 1-\alpha_i \right) F_S(t) - F_0(t)}{ -k_i/h_i - F_T(t)} \right] .
\label{eq:ice0}
\end{equation}
Here the dependence of $T_i$ on $t$ and $h_i$ has been explicitly indicated, and the ramp function $\mathcal{R}(x)$ is defined to be $\mathcal{R}(x)=0$ if $x<0$ and $\mathcal{R}(x)=x$ if $x\ge 0$. Note that the two surface boundary conditions in equation \eqref{eq:MU71top} are compactly embodied in the ramp function in equation \eqref{eq:ice0}. The thickness evolution in both cases \eqref{eq:T92-freeze2b} can be written after inserting equation \eqref{eq:Ftop} as
\begin{equation}
L_i\frac{d h_i}{dt}=-(1-\alpha_i) F_S(t) + F_0(t) + F_T(t) T_i(t,h_i)-F_B.
\label{eq:ice1}
\end{equation}

The sea ice model is fully contained in equations \eqref{eq:ice0}-\eqref{eq:ice1}. The results of this idealized ice thermodynamics model forced by specified surface and basal fluxes as in MU71 are shown in Fig.~S2 (blue curves), which indicates that this approximate representation yields results in good agreement with the full numerical solution to the partial differential equation \eqref{eq:MU71-T-si} in MU71 (cf. refs. \citen{Thorndike-1992:toy}, \citen{Semtner-1976:model}).

While most aspects of horizontal sea ice dynamics are neglected in this idealized treatment, in the coupled version of the model (equations {\bf [1]}-{\bf [4]} of the Research Report) we parameterize the net annual export of sea ice out of the central Arctic, most of which escapes through Fram Straight. Arctic sea ice has a residence time of roughly 3-12 years  \cite{Rigor-Wallace-2004:variations}, with a net annual export of about 10\% of the ice area \cite{Kwok-Cunningham-Pang-2004:fram}. This continuous export makes the ice thickness somewhat more stable: to maintain thicker ice, a larger amount of new ice must be produced each year. We approximately account for this by adding to the ice thickness evolution \eqref{eq:ice1} a decay term $-v_0 L_i h_i$, with $v_0=0.1$ yr$^{-1}$.

\section{Atmosphere}

In the presence of significantly different Arctic Ocean surface conditions, such as an exposed ocean mixed layer, the atmospheric energy fluxes into the surface are also expected to change significantly. This is particularly true for the downwelling longwave radiation which includes the effects of both horizontal atmospheric heat flux convergence and downward emission of absorbed upward longwave radiation due to the opacity of the atmosphere (i.e., the greenhouse effect). Here we use an idealized atmospheric model to account for changes in downwelling longwave radiation. This allows us to approximate $F_{top}(t,T,\alpha)$ over a wide range of climates. The derivation that follows is similar to previous treatments of two-stream radiative atmospheres (e.g., refs.\ \citen{Thorndike-1992:toy,Goody-Yung-1989:atmospheric, Salby-1996:fundamentals}).

\subsection{Heat Flux Convergence}

The meridional heat flux convergence averaged over 70\deg N--90\deg N is equivalent to a spatially averaged vertical flux of roughly $D=100$ \Wm{}  \cite{Nakamura-Oort-1988:atmospheric}. Since the poleward heat flux in the atmosphere is related to transport of sensible and latent heat by eddies, it is often approximated in idealized climate models as being proportional to the meridional temperature difference \cite{Chen-Gerdes-Lohmann-1995:d,Thorndike-1999:minimal,Gildor-Tziperman-2001:sea}, which is equivalent to assuming meridional effective diffusion of temperature as in typical atmospheric energy balance models \cite{Budyko-1969:effect,Sellers-1969:global,North-Cahalan-James-1981:energy}. Although a destabilizing increase in atmospheric meridional heat flux into the Arctic may occur in response to warming due to factors including increased humidity \cite{Alexeev-2003:sensitivity,Hall-2004:role,Winton-2006:amplified,Held-Soden-2006:robust}, if the warming is significant then reduced atmospheric heat transport is expected to be a principal damping mechanism \cite{Winton-2007:sea}. Here we follow the convention of setting the meridional heat flux to be proportional to the meridional temperature difference,
\begin{equation}
D(t,T)=k_D \Delta T_{merid}(t) ,
\label{eq:atm03}
\end{equation}
where $\Delta T_{merid}=T_{south}(t)-T$ with $T$ the simulated surface temperature in the Arctic and $T_{south}(t)$ the seasonally varying temperature south of the Arctic which is specified here from NCEP-NCAR reanalysis 1971-2000 climatological 1000mb atmospheric temperature \cite{Kalnay-Kanamitsu-Kistler-et-al-1996:ncep} spatially averaged from the equator to 70\deg N. We use $k_D=2.7$ \Wm/K, which optimizes the match to observed poleward heat transport \cite{Nakamura-Oort-1988:atmospheric} (although this parameterization leads to a model annual cycle in $D$ that is somewhat exaggerated compared to observations).

\subsection{Longwave Absorption}

We use a vertically continuous dry energy balance atmospheric model. We approximate there to be no absorption of shortwave radiation in the atmosphere and no scattering of longwave radiation. Longwave radiation is absorbed and emitted in continuous vertical levels with an absorption cross section that is independent of wavelength, temperature, and pressure, and the radiative fields are solved using a two-stream approximation. We assume that the poleward atmospheric heat transport into the Arctic, $D(t,T)$, is distributed uniformly in optical height \cite{Thorndike-1992:toy}.

The intensity of a beam of radiation propagating vertically upward from Earth's surface will diminish with height $z$ according to $dI/dz=-\rho(z)\kappa(z) I $, where $\rho(z)$ is the atmospheric density and $\kappa(z)$ is the extinction coefficient. This can be solved for intensity as a function of height,
\begin{equation}
I=I_0\exp\left[-\int_0^z \rho(z^\prime) \kappa(z^\prime) dz^\prime\right]=I_0\exp[-\tau(z)] ,
\end{equation}
where $I_0$ is the intensity at the surface and the optical height is $\tau(z)\equiv \int_0^z \rho(z^\prime) \kappa(z^\prime) dz^\prime$. We measure height using $\tau(z)$ instead of $z$, which has the advantage that $\kappa(z)$ and $\rho(z)$ are eliminated from the equations and the atmosphere can be approximately described by a single parameter, the total optical thickness $\tau_{\scriptscriptstyle{1}}\equiv \tau(z\to\infty)$. Note that our use of optical height differs slightly from the standard convention of using optical depth, which is integrated from the top of the atmosphere downward. Regarding the physical meaning of the optical thickness $\tau_{\scriptscriptstyle{1}}$, note that the fraction of longwave radiation emitted vertically from the surface that escapes to space is $\exp(-\tau_{\scriptscriptstyle{1}})$. A slanted optical path in the atmosphere, $\delta \tau^*$, can be related to optical height according to $\delta \tau=\delta \tau^* \cos \theta$, where $\theta$ is the angle the path makes with the vertical (Fig.~S3).

The intensity of longwave radiation in the atmosphere, $I(\tau,\theta,\phi)$, is a scalar field that depends on optical height and direction, with $\phi$ being the azimuthal angle (Fig.~S3). We model the atmosphere as a grey material that absorbs a fraction $\delta \tau^*$ of the intensity passing through it and emits an equivalent fraction $\delta \tau^*$ of its blackbody radiation. The blackbody radiation of an air parcel can be computed from the Stefan-Boltzmann equation, $B(T_a)=\sigma T_a^4 / \pi$, where $T_a$ is the temperature of the air parcel and the factor $\pi$ accounts for radiation occurring in every direction from a point source. This gives a change in intensity of $\delta I=-I \delta \tau^* + B \delta \tau^*$, which becomes the Schwarzschild equation when written in terms of optical height:
\begin{equation}
\cos\theta \frac{\partial I(\tau,\theta,\phi)}{\partial \tau}=-I(\tau,\theta,\phi)+B(\tau) .
\label{eq:schwartz}
\end{equation}
Note that $B$ is independent of angle since a blackbody emits radiation equally in all directions.

We assume horizontally uniform radiation from the surface and a horizontally homogenous atmospheric medium, which makes the intensity horizontally isotropic due to azimuthal symmetry, $I(\tau,\theta,\phi)=I(\tau,\theta)$. In thermal steady-state, this can be written as a vertically constant divergence of vertical net flux,
\begin{equation}
\frac{\partial}{\partial \tau}  \int d\omega ~ I(\tau,\theta) \cos \theta  = \frac{D(t,T)}{\tau_{\scriptscriptstyle{1}}} ,
\label{eq:d1}
\end{equation}
where we have defined the integral over all solid angles, $\int d\omega \equiv \int_{0}^{\pi} d\theta \int_{0}^{2\pi} \sin\theta d\phi$. We can rewrite the divergence condition \eqref{eq:d1} as an algebraic equation by equating it with the integral over all solid angles of the Schwarzschild equation \eqref{eq:schwartz}. Solving this for $B(\tau)$ and inserting this into the Schwarzschild equation \eqref{eq:schwartz} gives a single integro-differential equation for $I(\tau,\theta)$,
\begin{equation}
\cos\theta \frac{\partial I(\tau,\theta)}{\partial \tau}=-I(\tau,\theta)+\frac{1}{4\pi}\left[\frac{D(t,T)}{\tau_{\scriptscriptstyle{1}}}+ \int d\omega ~ I(\tau,\theta)\right]
\label{eq:schwartz2} .
\end{equation}

\begin{figure}
\centerline{\includegraphics[width=0.48\textwidth]{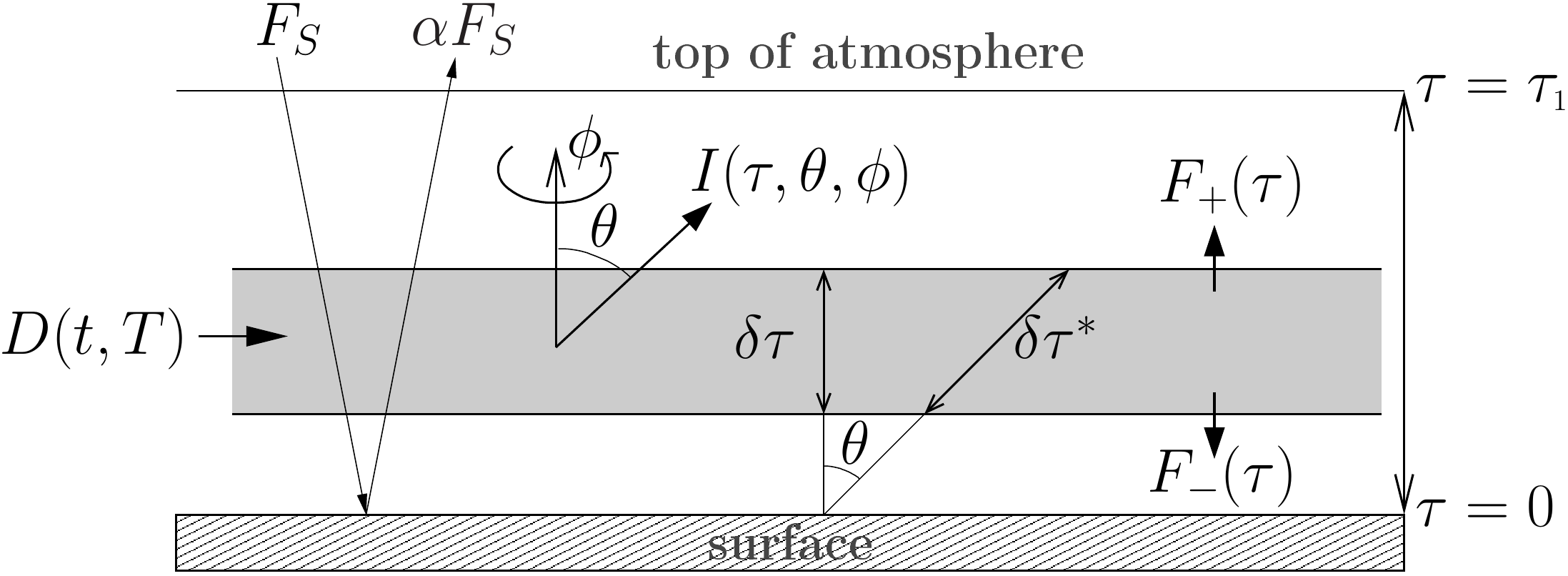}}
\caption{Schematic of atmospheric model for computing $F_{top}(t,T,\alpha)$. $D(t,T)$ represents meridional heat transport, and $(1-\alpha) F_S$ is the amount of absorbed solar radiation. Longwave radiative intensity is represented by $I$ and optical height is given as $\tau$ with $\delta \tau^*$ an optical path at angle $\theta$ to the vertical. Note that here the optical height increases upward, in contrast with the optical depth which increases downward and is typically used in radiative transfer calculations. The total upward and downward longwave radiative fluxes through a horizontal surface are $F_+(\tau)$ and $F_ -(\tau)$, respectively. The model allows the surface incident longwave radiation to be represented as a function of outgoing surface longwave radiation (equation \eqref{eq:atmFdn}).}
\label{fig:atm-schematic}
\end{figure}

The boundary conditions are that there is no downward longwave radiation at the top of the atmosphere ($\tau=\tau_{\scriptscriptstyle{1}}$),
\begin{equation}
I(\tau_{\scriptscriptstyle{1}},\pi/2 \le \theta \le \pi)=0 ,
\label{eq:atmbc1}
\end{equation}
and that the upward radiative flux from the surface ($\tau=0$) is $\sigma_0+\sigma_T T$,
\begin{equation}
I(0,0 \le \theta \le \pi/2)=\frac{\sigma_0+\sigma_T T}{\pi} .
\label{eq:atmbc2}
\end{equation}
The system \eqref{eq:schwartz2}-\eqref{eq:atmbc2} uniquely specifies atmospheric longwave radiation $I(\tau,\theta)$ given the poleward heat transport $D(t,T)$, atmospheric optical thickness $\tau_{\scriptscriptstyle{1}}$, and surface temperature $T$.

Here we use a standard two-stream approximation to arrive at an analytical solution to the system \eqref{eq:schwartz2}-\eqref{eq:atmbc2}. Multiplying the Schwarzschild equation \eqref{eq:schwartz} by $\cos\theta$ and integrating over the upper and lower hemispheres (i.e., the upper and lower halves of an infinitesimal sphere surrounding a given point in the atmosphere) leads to
\begin{equation}
\frac{\partial}{\partial \tau} \int d\omega^\uparrow ~ I(\tau,\theta) \cos\theta = -F^\uparrow(\tau)+\pi B(\tau) ,
\label{eq:schwartzup90}
\end{equation}
\begin{equation}
\frac{\partial}{\partial \tau} \int d\omega^\downarrow ~ I(\tau,\theta) \cos\theta = F^\downarrow(\tau)-\pi B(\tau) ,
\label{eq:schwartzdown0}
\end{equation}
where we have defined total upward and downward fluxes through a horizontal surface in the atmosphere, 
\begin{equation}
F^\uparrow(\tau) \equiv \int d\omega^\uparrow~I(\tau,\theta) \cos\theta ,
\label{eq:fup}
\end{equation}
\begin{equation}
F^\downarrow(\tau) \equiv -\int d\omega^\downarrow~I(\tau,\theta) \cos\theta ,
\label{eq:fdn}
\end{equation}
and the integrals over solid angles in each hemisphere are defined as $\int d\omega^\uparrow \equiv \int_{0}^{\pi/2} d\theta \int_{0}^{2\pi} \sin\theta d\phi$ and $\int d\omega^\downarrow \equiv \int_{\pi/2}^{\pi} d\theta \int_{0}^{2\pi} \sin\theta d\phi$.

The type of two-stream approximation we employ, an exponential kernel approximation (sec.~2.4 of ref.~\citen{Goody-Yung-1989:atmospheric}), is equivalent to assuming that radiation propagates through the atmosphere only at one effective angle. This angle deviates from the vertical to account for the fact that in the exact solution radiation propagates in all directions. Upwelling radiation propagates along $\theta=\theta_{\textrm{eff}}$ and downwelling radiation propagates along $\theta=\theta_{\textrm{eff}}+\pi$. This assumption allows equations \eqref{eq:schwartzup90}-\eqref{eq:schwartzdown0} to be written as
\begin{equation}
\cos\theta_{\textrm{eff}} \frac{\partial F^\uparrow(\tau)}{\partial \tau}  = -F^\uparrow(\tau)+\pi B(\tau) ,
\label{eq:schwartzup}
\end{equation}
\begin{equation}
\cos\theta_{\textrm{eff}} \frac{\partial F^\downarrow(\tau)}{\partial \tau}  = F^\downarrow(\tau)-\pi B(\tau) .
\label{eq:schwartzdown}
\end{equation}
Note that the negative sign from the definition of $F^\downarrow$ \eqref{eq:fdn} is cancelled in equation\eqref{eq:schwartzdown} because $\cos\left(\theta_\textrm{eff}+\pi\right)=-\cos\theta_\textrm{eff}$.

Inserting the definitions \eqref{eq:fup}-\eqref{eq:fdn} into the divergence condition \eqref{eq:d1} leads to\begin{equation}
\frac{\partial}{\partial \tau} \left[F^\uparrow(\tau)-F^\downarrow(\tau) \right]  = \frac{D(t,T)}{\tau_{\scriptscriptstyle{1}}} ,
\label{eq:d2}
\end{equation}
which can be multiplied by $\cos\theta_{\textrm{eff}}$ and then equated with the sum of equations \eqref{eq:schwartzup} and \eqref{eq:schwartzdown}. This allows us to write the divergence condition as an algebraic relation that can be directly solved for $B(\tau)$:
\begin{equation}
B(\tau)=\frac{1}{2 \pi}\left[\frac{D(t,T) \cos\theta_{\textrm{eff}}}{\tau_{\scriptscriptstyle{1}}}+F^\uparrow+F^\downarrow \right] .
\label{eq:d3}
\end{equation}

Finally, inserting the definitions \eqref{eq:fup}-\eqref{eq:fdn} into the boundary conditions \eqref{eq:atmbc1}-\eqref{eq:atmbc2} leads to
\begin{equation}
F^\downarrow(\tau_{\scriptscriptstyle{1}})=0 ,
\label{eq:atmbc3}
\end{equation}
\begin{equation}
F^\uparrow(0)=\sigma_0+\sigma_T T .
\label{eq:atmbc4}
\end{equation}

Equations \eqref{eq:schwartzup}-\eqref{eq:schwartzdown}, \eqref{eq:d3} represent a system of two coupled first order linear ordinary differential equations, which can be solved subject to boundary conditions \eqref{eq:atmbc3}-\eqref{eq:atmbc4} to give the full vertical profiles of $F^\uparrow(\tau)$ and $F^\downarrow(\tau)$. Note that the temperature profile can be calculated from this solution using equation \eqref{eq:d3} and the Stefan-Boltzmann equation. From the full solution to the system \eqref{eq:schwartzup}-\eqref{eq:schwartzdown}, \eqref{eq:d3}-\eqref{eq:atmbc4} (not shown), the surface downward longwave radiation is
\begin{equation}
F^\downarrow(0)=\frac{\sigma_0+\sigma_T T}{1+\frac{2\cos\theta_\textrm{eff}}{\tau_{\scriptscriptstyle{1}}}}+\frac{D(t,T)}{2} .
\label{eq:atmFdn}
\end{equation}
The incident surface radiation includes half of the atmospheric heat convergence, $D(t,T)/2$, while the other half is emitted to space. This arises from the assumption that $D(t,T)$ is evenly distributed in optical height.

The choice of $\theta_\textrm{eff}$ can be optimized to match the exact solution of the system \eqref{eq:schwartz2}-\eqref{eq:atmbc2} given values of $D(t,T)$, $\tau_1$, and $F^\uparrow(0)=\sigma_0+\sigma_T T$. Salby (sec.~8.4.2 of ref.~\citen{Salby-1996:fundamentals}) finds that the effect of averaging over spectral bands suggests the value $\cos\theta_\textrm{eff}=\frac{\scriptscriptstyle{3}}{\scriptscriptstyle{5}}$. Goody and Yung (sec.~9.2.1 of ref.~\citen{Goody-Yung-1989:atmospheric}) derive an atmosphere similar to the system \eqref{eq:schwartzup}-\eqref{eq:schwartzdown}, \eqref{eq:d3}, with $D(t,T)=0$, by using a version of the two-stream approximation in which hemispheric isotropy is assumed: $I(\tau,\theta>\pi/2)=I_+$ and $I(\tau,\theta>\pi/2)=I_-$. This yields a result equivalent to letting $\cos\theta_\textrm{eff}=\frac{\scriptscriptstyle{2}}{\scriptscriptstyle{3}}$. Thorndike \cite{Thorndike-1992:toy} derives an atmosphere analogous to the model derived here but with only vertically propagating radiation, arriving at a result equivalent to $\cos\theta_\textrm{eff}=1$. We take the optical thickness $\tau_1$ as the tunable parameter in the model, and we leave $\cos\theta_\textrm{eff}$ unspecified because it simply scales the optical thickness.

The net longwave radiation from the solution \eqref{eq:atmFdn} is
\begin{equation}
F^\uparrow(0)-F^\downarrow(0)=\kappa_{\scriptscriptstyle{LW}}\left(\sigma_0+\sigma_T T\right)-\frac{D(t,T)}{2} ,
\label{eq:atm01}
\end{equation}
where we have defined an atmospheric greenhouse factor as
\begin{equation}
\kappa_{\scriptscriptstyle{LW}} \equiv 1 - \frac{1}{1+\frac{2\cos\theta_\textrm{eff}}{\tau_{\scriptscriptstyle{1}}}}
=\frac{1}{1+\frac{\tau_1}{2 \cos\theta_\textrm{eff}}} .
\label{eq:atm02}
\end{equation}
Note that $0<\kappa_{\scriptscriptstyle{LW}}<1$.
This makes clear the effect of the atmospheric longwave radiation model: it mitigates surface longwave cooling by emitting some of the energy back to the surface, equivalent to reducing the surface upward longwave radiation by the factor $\kappa_{\scriptscriptstyle{LW}}$, and it adds energy associated with atmospheric heat flux convergence (i.e., net meridional heat transport into the Arctic). By reducing the surface net upward longwave radiation, the interactive atmospheric model weakens the stabilizing influence of outgoing longwave radiation on the ice/ocean system.

Grey two-stream atmospheres like the model used here can capture many of the basic features of radiative transfer in an approximate way. A thorough comparison of various types of two-stream approximations is discussed in Goody and Yung (sec.~2.4 of ref.~\citen{Goody-Yung-1989:atmospheric}).

The optical thickness of the atmosphere depends on water vapor, cloud particles, and greenhouse gases such as carbon dioxide. It is higher during summer than during winter because of increased water vapor and cloudiness. We specify the optical thickness seasonal cycle to follow observed Arctic cloudiness,
\begin{equation}
\kappa_{\scriptscriptstyle{LW}}(t) = \frac{1}{\tau_0+\tau_c f_{c}(t)},
\label{eq:atm04}
\end{equation}
where $f_{c}(t)$ is the Arctic cloud fraction seasonal cycle specified from observations \cite{Maykut-Church-1973:radiation}
and $\tau_0$ and $\tau_c$ are chosen to give a sea ice seasonal cycle matching that computed using forcing from MU71 (cf. refs. \citen{Thorndike-1992:toy}, \citen{Bjork-Soderkvist-2002:dependence}).
This leads to a choice of $\tau_0=0.5$ and $\tau_c=3.6$.

The actual energy flux at the top of the sea ice or exposed ocean mixed layer, $F_{top}(t,T,\alpha)$, includes components of sensible and latent heat fluxes in addition to downward and upward shortwave and longwave radiation. According to the observationally-based central Arctic values specified in MU71, the sensible and latent heat fluxes are small compared to the radiative fluxes, and here we effectively approximate the sensible and latent heat fluxes by incorporating them into the computed downwelling longwave flux. The longwave emissivities of ice and open water, both roughly 0.95--1, are here approximated to unity. Under these approximations, the total surface flux can be written
\begin{equation}
F_{top}(t,T,\alpha)=\kappa_{{\scriptscriptstyle{LW}}}(t)\left(\sigma_0+\sigma_T T\right)-\frac{D(t,T)}{2} -(1-\alpha)F_S(t) ,
\label{eq:atm4b}
\end{equation}
where the downwelling shortwave radiation incident at the surface, $F_S(t)$, is specified from observations as in MU71. Inserting equation \eqref{eq:atm03}, we see that the full temperature dependence is linear, allowing us to write $F_{top}$ \eqref{eq:atm4b} in the form of equation \eqref{eq:Ftop} with parameters
\begin{equation}
F_0(t)= \kappa_{{\scriptscriptstyle{LW}}}(t) \sigma_0-\frac{k_D}{2} T_{south}(t)
\label{eq:atm05}
\end{equation}
and
\begin{equation}
F_T(t)=\kappa_{{\scriptscriptstyle{LW}}}(t)\sigma_T+\frac{k_D}{2} .
\label{eq:atm06}
\end{equation}

\section{Ocean Mixed Layer}

To allow the simulation of ice-free conditions, we include a representation of an ocean mixed layer which becomes exposed when all of the ice ablates. The mixed layer is represented as a thermodynamic reservoir with a characteristic depth of $H_{ml}=50$m, in agreement with observations \cite{Morison-Smith-1981:seasonal}. The mixed layer temperature evolution is proportional to the net flux,
\begin{equation}
c_{ml} H_{ml} \frac{d T_{ml}}{dt}= \left( 1-\alpha_{ml} \right) F_S(t) -F_0(t)- F_T(t) T_{ml} +F_B ,
\label{eq:ocean1}
\end{equation}
with mixed layer heat capacity $c_{ml}=4\times10^6$ Jm$^{-3}$K$^{-1}$. We use an open water albedo of $\alpha_{ml}=0.2$, similar to previous studies \cite{Thorndike-1992:toy, Hibler-1979:dynamic}, to account for the presence of small amounts of thin ice in a largely ice-free
Arctic Ocean. When the ice completely melts ($h_i=0$), the ocean mixed layer temperature is evolved forward from $T_{ml}=0$, and when the mixed layer cools back to $T_{ml}=0$, the ice thickness is evolved once again starting from $h_i=0$.

\section{Coupled Model}

The separate equations for $T_{ml}$ and $h_i$ can be combined, since only one is evolving at any given time. We define the energy per unit area in the system, $E$, to be equal to the sum of the latent heat content of the sea ice and the specific heat content of the ocean mixed layer (equation {\bf [1]} in the Research Report).
This allows the ice and ocean mixed layer components of the idealized model \eqref{eq:ice0}-\eqref{eq:ice1}, \eqref{eq:ocean1} to be expressed as equations {\bf [2]}-{\bf [3]} in the Research Report. The parameters $F_0(t)$ and $F_T(t)$, which are used to determine the surface energy flux, have values computed using the atmospheric model \eqref{eq:atm05}-\eqref{eq:atm06}. An imposed annually constant surface energy flux is included in equations {\bf [2]}-{\bf [3]} by replacing $F_0(t)$ with $F_0(t)-\Delta F_0$. Values for the parameters in equations {\bf [1]}-{\bf [4]} in the Research Report are given in Table S1. The standard case simulation with this model, illustrated in Fig.~S2, produces central Arctic sea ice conditions in fairly good agreement with MU71.

\renewcommand{\thefigure}{S1}
\renewcommand{\figurename}{Table}
\begin{table*}[h]
\noindent {\begin{justify} {\fignumfont Table S1.} \small \sf Descriptions and default values of model parameters. Time evolution $t$ is measured in years while fluxes are measured in Wm$^{-2}$, which allows most dimensional parameters to be approximately of order unity but requires a non-conventional choice of units for energy per unit area $E$ (written in Wm$^{-2}$yr), heat capacity $c_{ml} H_{ml}$, and latent heat $L_i$. For the three seasonally varying parameters, the annual mean value is given in the table; the monthly values starting with January are $F_0(t)$ = (120, 120, 130, 94, 64, 61, 57, 54, 56, 64, 82, 110) Wm$^{-2}$, $F_T(t)$ = (3.1, 3.2, 3.3, 2.9, 2.6, 2.6, 2.6, 2.5, 2.5, 2.6, 2.7, 3.1) Wm$^{-2}$K$^{-1}$, and $F_S(t)$ = (0, 0, 30, 160, 280, 310, 220, 140, 59, 6.4, 0, 0) Wm$^{-2}$. \end{justify}}
\begin{tabular*}{1.\textwidth}{@{\extracolsep{\fill}} lll}
\hline
\textbf{Symbol} & \textbf{Description} & \textbf{Value}\\
\hline
$L_i$ & Latent heat of fusion of ice & 9.5 Wm$^{-3}$yr \\
$c_{ml} H_{ml}$ & Ocean mixed layer heat capacity times depth & 6.3 Wm$^{-2}$yrK$^{-1}$ \\
$\alpha_i$ & Albedo when surface is ice-covered & 0.68 \\
$\alpha_{ml}$ & Albedo when ocean mixed layer is exposed & 0.2 \\
$k_i$ & Ice thermal conductivity & 2 Wm$^{-1}$K$^{-1}$ \\
$F_B$ & Heat flux into bottom of sea ice or ocean mixed layer & 2 Wm$^{-2}$ \\
$h_{\alpha}$ & Ice thickness range for smooth transition from $\alpha_i$ to $\alpha_{ml}$ & 0.5 m\\
$v_0$ & Dynamic export of ice from model domain & 0.1 yr$^{-1}$ \\
$F_0(t)$ & Temperature-independent surface flux (seasonally varying) & 85 Wm$^{-2}$\\
$F_T(t)$ & Temperature-dependent surface flux (seasonally varying) & 2.8 Wm$^{-1}$K$^{-1}$ \\
$F_S(t)$ & Incident shortwave radiation flux (seasonally varying) & 100 Wm$^{-2}$ \\
$\Delta F_0$ & Imposed surface heat flux & 0 Wm$^{-2}$ \\
\hline
\end{tabular*}
\end{table*}

\bibliographystyle{pnas}

\endtwocolumns\vskip110pt\twocolumns

\setcounter{figure}{3}
\renewcommand{\figurename}{Fig.}
\renewcommand{\thefigure}{S\arabic{figure}}

\begin{figure}
\centerline{\includegraphics[width=0.27\textwidth]{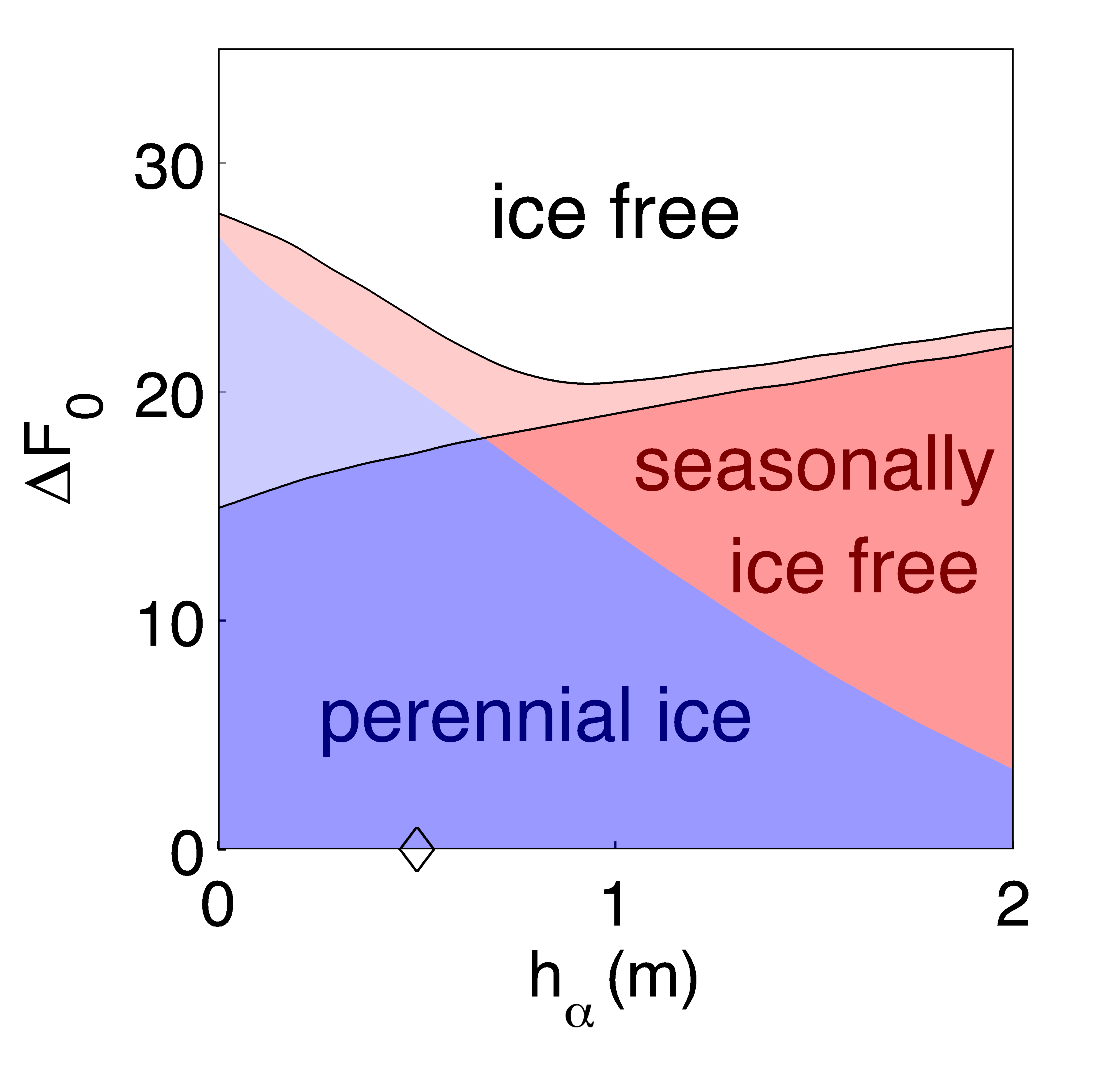}}
  \caption{Robustness of the results in Fig.~3 of the Research Report to parameter regime, illustrated here by varying the parameter governing the smoothness of the albedo transition ($h_\alpha$). For each value of $h_\alpha$, ranges of surface heating ($\Delta F_0$) that give rise to stable solutions that are perennially ice-covered (blue region), seasonally ice-free (red region), or perennially ice-free (white region) are identified. The default parameter regime is indicated by a black diamond. Mixed shades indicate the overlap in regions where multiple stable solutions coexist, and the bifurcation curves marking the edges of this space are indicated by black curves. The lack of any purple region, which would indicate an overlap between red (seasonally ice-free) and blue (perennial ice), demonstrates that multiple states are not found with the warm state being seasonally ice-free, while the presence of light red and light blue regions shows that multiple states with the warm state being perennially ice-free are possible. The variation of other model parameters (not shown) leads to similar results. This indicates that although the size of the $\Delta F_0$ range where multiple solutions coexist depends on $h_\alpha$, both {\it (i)} the lack of a bifurcation threshold during the transition from perennial ice to seasonally ice-free conditions and {\it (ii)} the presence of multiple states and threshold behavior during the transition to perennially ice-free conditions are robust features of the model equations.}
\label{fig:bif-curve}
\end{figure}

\begin{figure}
\centerline{\includegraphics[width=0.35\textwidth]{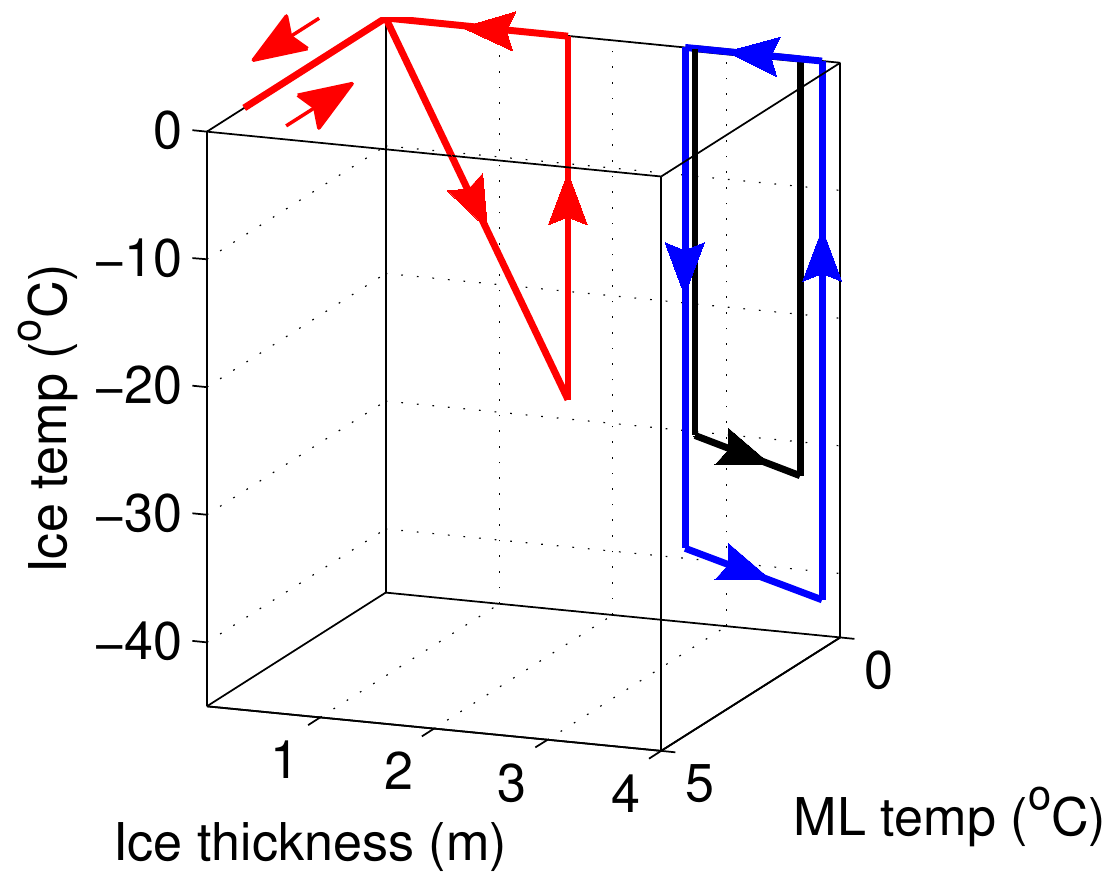}}
  \caption{Seasonal cycle of Arctic sea ice and ocean conditions simulated with the toy model of Thorndike \cite{Thorndike-1992:toy}. Seasonally varying solutions are plotted as closed curves in the three-dimensional model state space, which represents changes in sea ice thickness, sea ice surface temperature, and ocean mixed layer temperature. The standard case solution is indicated by the black curve. Thorndike found that  when the  specified atmospheric Arctic heat flux convergence $D$ was increased,   the model transitioned from perennially ice-covered to perennially ice-free conditions, with no stable seasonally ice-free solution possible. Rather than prescribing observed seasonally-varying forcing quantities, Thorndike assumed a step-function form for the forcing, with shortwave radiation and optical thickness taking on constant values during the summer and winter half-years.  He found observationally consistent ice thickness with summer and winter optical thicknesses of 4.5 and 3, respectively (black curve). 
When we choose for these parameters instead 1.5 and 5, respectively, Thorndike's toy model simulates a relatively consistent 
approximation of the modern sea ice seasonal cycle (blue curve). Increasing $D$ from 100 Wm$^{-2}$ to 145 Wm$^{-2}$ in this regime, however, produces a stable seasonally ice-free solution (red curve), in contrast to the results reported by Thorndike. A second stable state which is perennially ice-free exists for the solution indicated by the black curve, as discussed by Thorndike, but not for the solution shown here by the blue curve. A second stable state which is perennially ice-free does however exist for the solution indicated by the red curve. The coexistence in Thorndike's model of a stable seasonally ice-free solution and a stable perennially ice-free solution is consistent with the results presented here (Fig.~3 of the Research Report).}
\end{figure}

\end{article}